\documentclass[11pt,twocolumn,twoside]{article}
\usepackage{fully3d}

\usepackage{amsmath}
\usepackage{amssymb}
\DeclareMathOperator*{\argmin}{arg\,min}
\usepackage{pifont}
\newcommand{\cmark}{\ding{51}}%
\newcommand{\xmark}{\ding{55}}%
\usepackage[shortcuts]{extdash}
\usepackage{siunitx}
\usepackage{threeparttable}
\usepackage{caption}
\usepackage{subcaption}

\begin{document}

\title{Optimizing CT Scan Geometries With and Without Gradients} 

\author[1]{Mareike~Thies}
\author[1]{Fabian~Wagner}
\author[1]{Noah~Maul}
\author[1]{Laura~Pfaff}
\author[1,2]{Linda-Sophie~Schneider}
\author[2]{Christopher~Syben}
\author[1]{Andreas~Maier}

\affil[1]{Pattern Recognition Lab,
          Friedrich-Alexander-Universit\"at Erlangen-N\"urnberg, Erlangen, Germany}
\affil[2]{Fraunhofer Development Center X-Ray Technology EZRT, Erlangen, Germany}

\maketitle
\thispagestyle{fancy}


\begin{customabstract}
In computed tomography (CT), the projection geometry used for data acquisition needs to be known precisely to obtain a clear reconstructed image. Rigid patient motion is a cause for misalignment between measured data and employed geometry. Commonly, such motion is compensated by solving an optimization problem that, e.g., maximizes the quality of the reconstructed image with respect to the projection geometry. So far, gradient-free optimization algorithms have been utilized to find the solution for this problem. Here, we show that gradient-based optimization algorithms are a possible alternative and compare the performance to their gradient-free counterparts on a benchmark motion compensation problem. Gradient-based algorithms converge substantially faster while being comparable to gradient-free algorithms in terms of capture range and robustness to the number of free parameters. Hence, gradient-based optimization is a viable alternative for the given type of problems.    
\end{customabstract}

\section{Introduction}
Numerous problems in computed tomography (CT) require optimizing the CT acquisition geometry based on a target function formulated on the reconstructed image. One of the most prominent examples is rigid motion compensation where the acquisition geometry is updated to compensate for involuntary patient motion occurring during the scan. This can be achieved by formulating target functions which quantify the quality of the reconstructed image via measures such as image entropy, total variation, or gradient entropy \cite{kingston2011, sisniega2017, preuhs2020, capostagno2021}. Minimization of the image quality criterion yields the motion parameters which produce the best reconstructed image according to the target function. 

So far, these approaches have been limited to gradient-free optimization algorithms because the parameters being optimized and the target function live in different domains which are connected via the reconstruction operator. The free parameters define the geometrical relationship between scanned object, X-ray source, and detector pixels. Meanwhile, the target function is formulated in image space. The image space depends on the scan geometry in a complex manner where a change in a single geometry parameter can influence the entire reconstructed image. As a result, formulating the gradient of a target function in image space with respect to the scan geometry is not trivial and needs to incorporate the reconstruction step. 

Recently, we proposed an algorithm for fan-beam CT geometries which computes all partial derivatives of the gray values in a reconstructed image with respect to the entries of the projection matrices parameterizing the scan geometry in CT \cite{thies2022}. These computations bridge the gap between image domain and geometry space and enable the formulation of gradients for the motion compensation problem mentioned above. Consequently, gradient-based optimization algorithms can be applied.  

In this paper, we investigate the performance of different gradient-free and gradient-based optimization algorithms on the same geometry optimization problem concerning run time, capture range, and robustness to the number of free parameters. Doing so, we analyse whether gradient-free algorithms are naturally a better choice for the given type of problems or if gradient-based optimization algorithms are a comparable or even superior alternative. 

\section{Methods}
\subsection{Analytical Geometry Gradients}
In our recent work \cite{thies2022}, we presented the mathematical derivation and implementation to compute the partial derivatives of the gray values in a reconstructed CT image with respect to the entries of the $2\times3$-shaped CT projection matrices in fan-beam geometry. Projection matrices are a common parameterization of the CT scan geometry and describe the geometrical relationship between a point in the reconstructed image and the detector coordinate onto which this point is mapped. This fully specifies the orientation of the object (extrinsic information) as well as the detector itself (intrinsic information). We refer the reader to \cite{thies2022} for a detailed explanation of the mathematical steps involved in the gradient computation. For this paper, we assume that all partial derivatives of the reconstructed image $I \in \mathbb{R}^{N_x \times N_y}$ with respect to all entries of the projection matrices $P \in \mathbb{R}^{N_p \times 2 \times 3}$ can be computed analytically, i.e., the partial derivative $\frac{\partial I}{\partial P}$ is given. Here, $N_x$ and $N_y$ are the image dimensions and $N_p$ is the number of projections. Computing $\frac{\partial I}{\partial P}$ is the crucial step for formulating the gradient of a target function $f: N_x \times N_y \rightarrow \mathbb{R}$ with respect to the $N_f$ free parameters $g \in \mathbb{R}^{N_f}$ influencing the scan geometry. This is because the derivatives $\frac{\partial f}{\partial I}$ and $\frac{\partial P}{\partial g}$ are usually straight-forward to formulate or can even be computed via automatic differentiation in common deep learning libraries. Together, these variables define the gradient of the target function $f$ with respect to the free parameters $g$ via the chain rule of differentiation
\begin{equation}
    \label{eq:1}
    \frac{\partial f}{\partial g} = \frac{\partial f}{\partial I} \cdot \frac{\partial I}{\partial P} \cdot \frac{\partial P}{\partial g} \enspace .
\end{equation}

\subsection{Optimization Algorithms}
\label{sec:algos}
We are interested in finding the optimal motion parameters $g^* \in \mathbb{R}^{N_f}$ by minimizing the target function $f$
\begin{equation}
    \label{eq:2}
    g^* = \argmin_{g \in \mathbb{R}^{N_f}} f(g) \enspace .
\end{equation}
In general, Eq.~\ref{eq:2} is an unconstrained, non-convex problem which, however, can be locally convex depending on the parameterization and initialization of $g$. Several numerical optimization algorithms exist to solve Eq.~\ref{eq:2} of which we compare the following ones: Covariance matrix adaptation evolution strategy (CMA\=/ES), Nelder\=/Mead Simplex, gradient descent, and Broyden\=/Fletcher\=/Goldfarb\=/Shanno (BFGS).

CMA\=/ES and Nelder\=/Mead Simplex are gradient-free algorithms which operate by repeatedly evaluating the target function until a minimum is met. Being an evolutionary algorithm, CMA\=/ES iteratively estimates a covariance matrix based on which new candidate solutions are generated. The Nelder\=/Mead Simplex algorithm evaluates the target function at the vertices of a $N_f + 1$ dimensional simplex followed by a number of possible operations on the simplex vertices designed to progress downhill on the target function. Both have successfully been applied to CT motion compensation via image quality criteria \cite{sisniega2017, preuhs2020, capostagno2021}.

Gradient-based optimization has not been applied extensively to the considered problem even though gradient information is known to aid minimization by providing the steepest descent direction at a certain point $\hat{g}$ \cite{bacher2021}. Gradient descent with a predefined step size updates the current solution by a step into the direction of the negative gradient multiplied by the step size. BFGS additionally uses an approximation of second order Hessian information and performs a line search along the descent direction instead of using a predefined step size.      

\subsection{Rigid Motion Compensation}
The extrinsic component of each projection matrix is updated to compensate for rigid, inter-frame patient motion. The search space of this problem consists of the three rigid motion parameters rotation ($r$) and translation ($t_x$, $t_y$) for each of the $N_p$ projections leading to a total of $3N_p$ free parameters. To reduce the dimensionality of the search space and limit it to realistic motion patterns, further constraints can be enforced on each of the parameters by means of a motion model $m: \mathbb{R}^{N_f} \times \mathbb{R}^{N_p \times 2 \times 3} \rightarrow \mathbb{R}^{N_p \times 2 \times 3}$ which takes free parameters $g \in \mathbb{R}^{N_f}$ and the current estimate of projection matrices $P_{n}$ to yield updated projection matrices $P_{n+1}$ 
\begin{equation}
    P_{n+1} = m(g, P_n) \enspace .
\end{equation}
In this case, the number of free parameters can be considerably smaller, i.e., $N_f \ll 3N_p$ \cite{preuhs2020}.

\section{Experiments}
\label{sec:experiments}
\subsection{Data}
\label{sec:data}
We simulate motion-affected fan-beam sinograms from real reconstructed CT slices of publicly available cone-beam CT data of the head \cite{moen2021}. The images are of size $512 \times 512$ and the pixel spacing is assumed to be \SI{1}{\milli\meter}. Fan-beam projection data are simulated for 360 projections over a full circle, a source-to-isocenter distance of \SI{1000}{\milli\meter}, a source-to-detector distance of \SI{2000}{\milli\meter}, and 1024 detector pixels with a spacing of \SI{2}{\milli\meter}. Forward projections are performed using the implementation in \cite{syben2019}. Comparable to previous work in \cite{sisniega2017}, the artificially introduced motion pattern is a step-like function for $r$, $t_x$, and $t_y$. From a start projection $p_{start}$, the perturbation increases linearly until it reaches its maximal amplitude at projection $p_{end}$ and stays constant for the rest of the scan. The maximal motion is \SI{\pm 10}{\milli\meter} for translation in x and y and $\pm 5.73^{\circ}$ ($\pm 0.1\,\text{rad}$) for rotation unless stated otherwise. $p_{start}$ and $p_{end}$ are chosen such that the motion pattern extends over 50 to 200 projections and is completed within the full 360 projections. The perturbation is only applied to the projection matrices, the projection data itself represent a circular motion-free trajectory.      

\subsection{Optimization Problem}
\label{sec:problem}
Starting from the perturbed projection matrices (Sec.~\ref{sec:data}), we aim to find the motion parameters $t_x$, $t_y$, and $r$ which optimally annihilate the introduced step-like motion pattern and restore a circular trajectory. This is done by minimizing the mean squared error (MSE) between the motion-affected reconstruction and the ground-truth, motion-free reconstruction with respect to the motion parameters. The utilized motion model $m$ enforces a smooth curve for each of the three motion parameters by fitting a cubic spline with $N_n$ nodes which are equally distributed across the full scan range. Hence, the number of free parameters is $N_f = 3N_n$, i.e., three times the number of nodes in the respective splines. The number of nodes is $N_n = 10$ if not specified differently. 

\begin{figure*}
     \centering
     \begin{subfigure}[b]{0.33\textwidth}
         \centering
         \includegraphics[width=\textwidth]{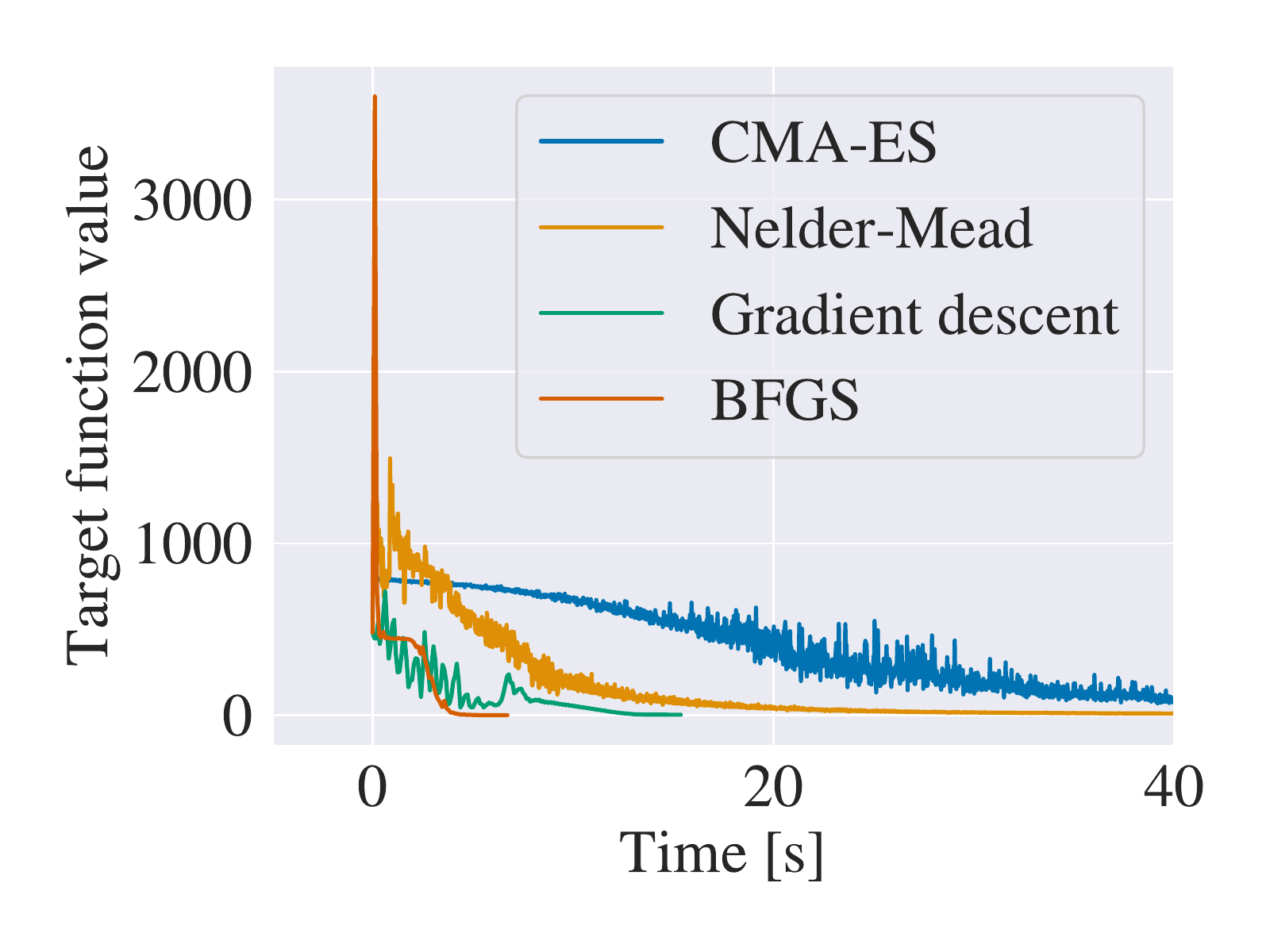}
     \end{subfigure}
     \hfill
     \begin{subfigure}[b]{0.33\textwidth}
         \centering
         \includegraphics[width=\textwidth]{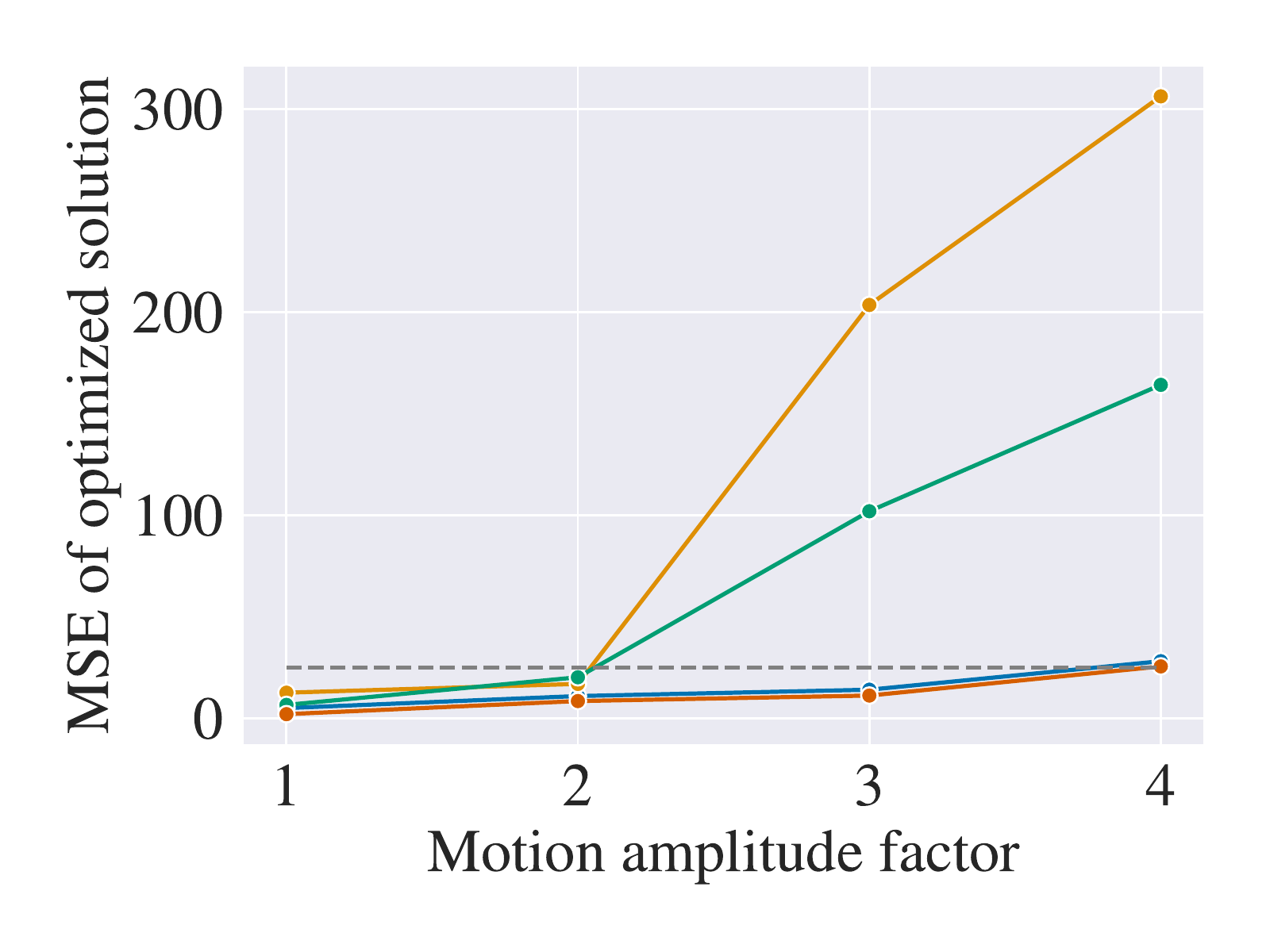}
     \end{subfigure}
     \hfill
     \begin{subfigure}[b]{0.33\textwidth}
         \centering
         \includegraphics[width=\textwidth]{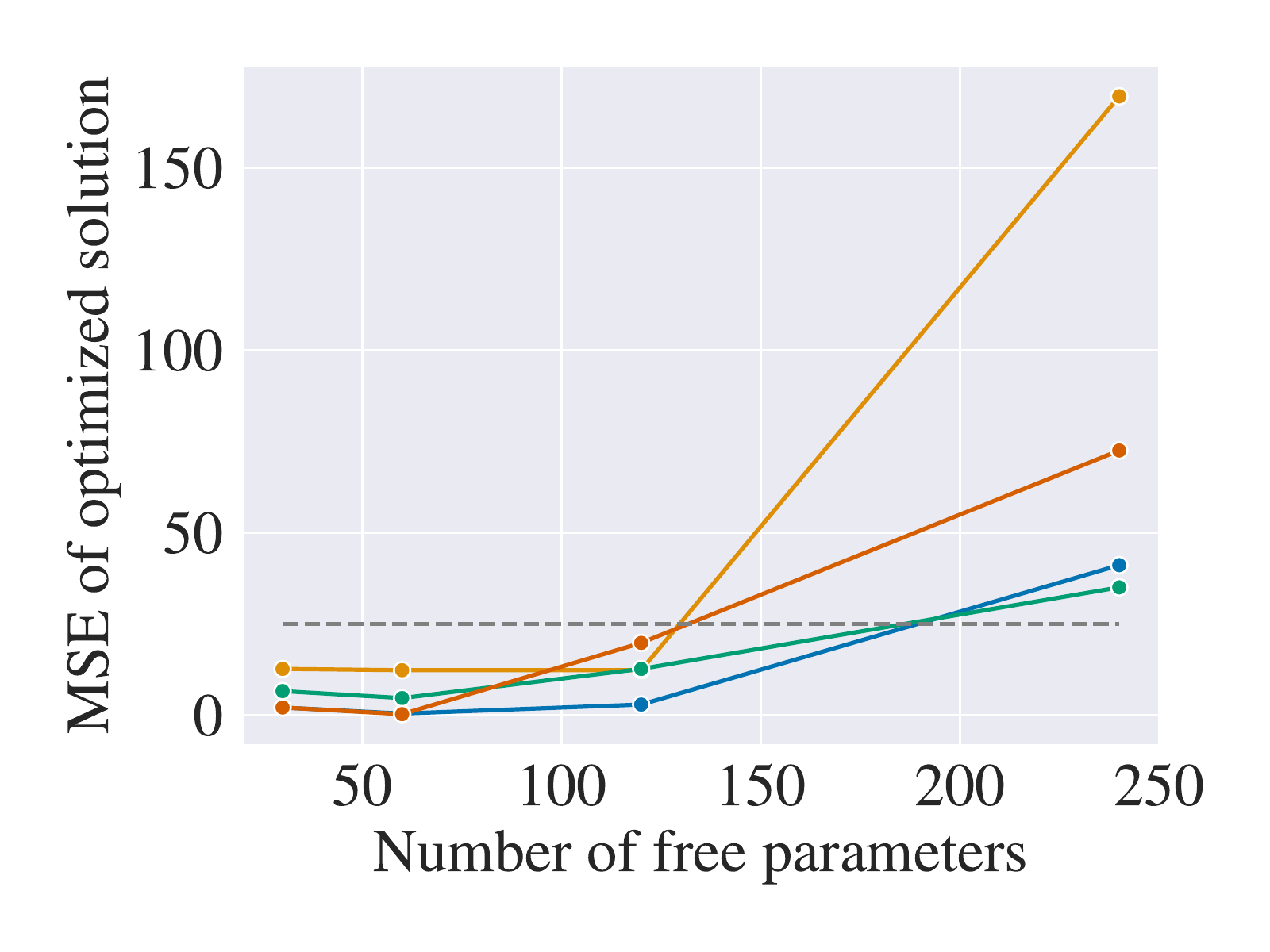}
     \end{subfigure}
        \vspace{-2\baselineskip}
        \caption{\textbf{Left}: Minimization of the MSE target function over time. Gradient-based optimization algorithms converge considerably faster than gradient-free algorithms. The plot does not show the gradient-free curves until full convergence. \textbf{Middle}: Evaluation of capture range. CMA-ES and BFGS converge successfully even for large motion perturbations. \textbf{Right}: Robustness against increase in free parameters. CMA-ES and gradient descent are most robust up to $\sim200$ free parameters. The gray dashed line marks an MSE of 25 below which we consider an optimization result successful. }
        \label{fig:lineplots}
\end{figure*}

\begin{table*}
\footnotesize
\centering
    \begin{threeparttable}
        \centering
            \begin{tabular}{ @{} l c l l l l l @{} }
            \hline
                                      & Gradient? & Source   & NFEV      & NJEV    & MSE    & Time [s]      \\ 
            \hline
            CMA-ES                    & \xmark & pycma$^1$  & $7330.84 \pm 761.61$    & $0.0$       & $0.99 \pm 0.52$ & $209.36 \pm 20.78$ \\  
            Nelder-Mead               & \xmark & scipy$^2$  & $3481.12 \pm 1751.20$   & $0.0$       & $6.28 \pm 2.95$ & $101.44 \pm 51.11$ \\
            Gradient descent          & \cmark & custom     & $670.68 \pm 1945.85$    & $670.68 \pm 1945.85$  & $4.16 \pm 1.46$ & $36.83 \pm 52.09$ \\
            BFGS                      & \cmark & scipy$^2$  & $69.56 \pm 14.46$       & $67.84 \pm 11.21$     & $1.21 \pm 0.65$ & $6.56 \pm 1.32$ \\
            \hline
            \end{tabular}
        \begin{tablenotes}
        \footnotesize
        \item $^1$\url{https://github.com/CMA-ES/pycma} \qquad $^2$\url{https://github.com/scipy/scipy}
        \end{tablenotes}
    \end{threeparttable}
\caption{Overview of compared optimization algorithms and quantitative results. We report the number of objective function evaluations (NFEV) and the number of gradient evaluations (NJEV) needed until convergence. Further, the target function value (MSE) of the final optimized solution and the time needed for convergence is given. Standard deviations are computed across five different anatomies and five different initial motion perturbations.}
\label{tab:table}
\end{table*}

\subsection{Optimizer Configurations}
The problem described in Sec.~\ref{sec:problem} is optimized using each of the four optimization algorithms introduced in Sec.~\ref{sec:algos}. All compared optimization algorithms are based on the same implementation of the target function and, if applicable, its gradient. As \cite{thies2022} is already implemented as a differentiable \textit{PyTorch} operator, we further use the deep learning library to implement the spline-based motion model\footnote{\url{https://github.com/patrick-kidger/torchcubicspline}} and the MSE target function. This way, the gradient of the loss function (Eq.~\ref{eq:1}) is obtained easily by means of automatic backpropagation. Both gradient-free optimization algorithms terminate as soon as the target function value changes by less than 0.1 or after 20000 function evaluations. The gradient-based optimization algorithms terminate once the gradient norm is less than 2 or after 500 iterations (BFGS)/ 10000 iterations (gradient descent). Initial standard deviations (CMA-ES) and step sizes (gradient descent) are adjusted to account for the different scales of translations [mm] and rotations [rad]. The step size in gradient descent decays by a factor of 0.995 after each iteration. Implementations of the optimization algorithms are summarized in Tab.~\ref{tab:table}.

\begin{figure*}
    \centering
    \includegraphics[width=\textwidth]{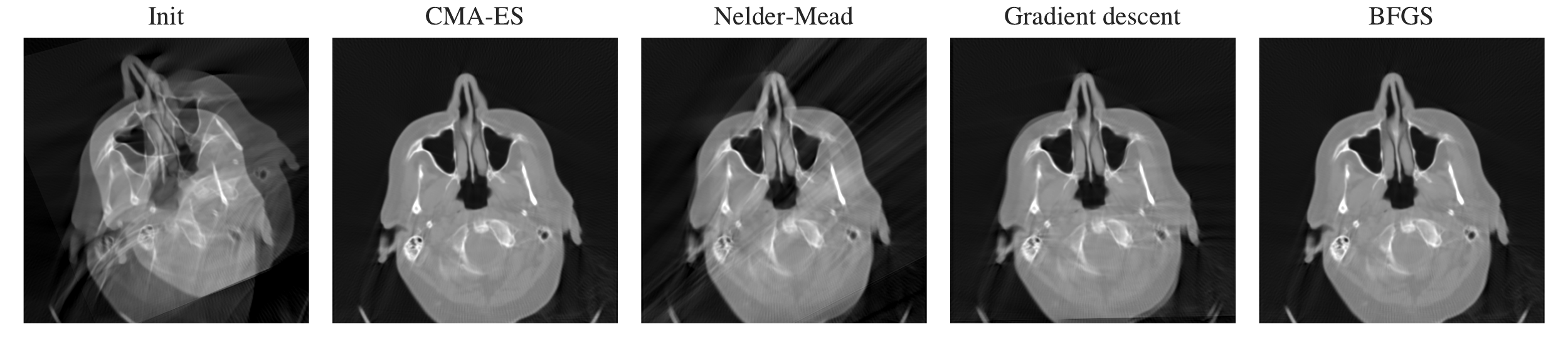}
    \caption{Reconstructed images recovered from the largest investigated motion amplitudes. The initial motion perturbation degrades the image severely. All optimization algorithms recover the shape of the head, but artifacts remain for Nelder-Mead and gradient descent.}
    \label{fig:recos}
\end{figure*}

\section{Results}
We first optimize the given problem with all four optimization algorithms using the standard settings described in Sec.~\ref{sec:experiments}. Each optimization algorithm is run 25 times using slices from five different patients with five realizations of the motion pattern each. Motion patterns vary concerning $p_{start}$ and $p_{end}$ as well as the direction of rotation and translations (positive or negative), but have the same amplitude. Each initialization and patient slice is used for all optimization algorithms identically. The resulting measurements are summarized in Tab.~\ref{tab:table}. The gradient-free optimization algorithms (CMA-ES and Nelder-Mead) have a considerably higher number of objective function evaluations (NFEV) than the gradient-based optimization algorithms (gradient descent and BFGS). For example, NFEVs for CMA-ES are two orders of magnitude higher than for BFGS. Both gradient-free optimization algorithms do not evaluate the gradient of the objective function and, hence, the number of gradient evaluations (NJEV) is zero whereas for the gradient-based algorithms NJEV $\approx$ NFEV. The MSE of the optimized solution to the ground truth is smallest for CMA-ES and BFGS. However, by visual inspection, we find that all MSE values below 25 represent successfully compensated images. The time until convergence exhibits the most notable differences. All gradient-based algorithms converge strikingly faster than the gradient-free counterparts, e.g., BFGS is over 30 times faster than CMA-ES for the investigated problem setting. This relationship is visible in the plot of the target function value over the optimization time in Fig.~\ref{fig:lineplots} (left) as well. \newline
To analyse the capture range of the optimization algorithms, we measure their performance for motion patterns with increasing amplitude. On the same image slice, the amplitude of the initial motion perturbation is increased from the default value of $\pm$\SI{10}{\milli\meter} to $\pm$\SI{40}{\milli\meter} for translations and from $\pm5.73^{\circ}$ to $\pm22.92^{\circ}$ for rotation. The results are summarized in Fig.~\ref{fig:lineplots} (middle), where the motion amplitude factor describes the multiplicative increase of the motion amplitude compared to the default configuration. Hence, the higher the motion amplitude factor, the higher is the distance between initialization and true solution of the optimization problem. All four optimization algorithms exhibit a stable behavior up to motion amplitude factor of 2, and CMA-ES and BFGS even succeed for factors as high as 4. Initial and recovered reconstructions from the largest investigated motion amplitude are depicted in Fig.~\ref{fig:recos}. Apparently, the initial motion degrades the image severely. Whereas CMA-ES and BFGS compensate the motion well, streaking artifacts remain for Nelder-Mead and the result optimized with gradient descent still exhibits slightly misaligned edges. 

Finally, we investigate the robustness of the four candidates to an increase in the number of spline nodes, and hence the number of free parameters, in the motion model. On the same image slice, we keep the motion pattern constant and vary the number of nodes per spline from 10 to 20, 40, and 80 leading to 30, 60, 120, and 240 free parameters (see Fig.~\ref{fig:lineplots} (right)). All algorithms succeed for up to 120 free parameters, but CMA-ES and gradient descent behave more stably for an even higher number of free parameters. 

\section{Discussion and Conclusion}
We compare gradient-free and gradient-based optimization algorithms for a geometry optimization problem in fan-beam CT. Gradient-based optimization algorithms converge substantially faster mainly due to a largely reduced number of target function evaluations. In the investigated case, where each target function evaluation is expensive as it incorporates a reconstruction step, a reduction in NFEV is essential to speed up the optimization. We acknowledge that our implementation is not optimized for run time, but it is comparable and, therefore, applicable for a relative comparison between the investigated methods. While it is known that gradient-based optimization algorithms are susceptible to terminating in local minima, our investigations concerning capture range reveal a behavior comparable to the gradient-free algorithms for the given, generally non-convex problem setting. In particular, BFGS converges to a solution close to the global minimum even for largely perturbed initializations. The same holds for the robustness with respect to the number of free parameters. In this case, gradient descent performs on par with CMA-ES and succeeds even for high-dimensional optimization problems. Note that the gradient-based algorithms converge still substantially faster in both these experiments. Of course, we further acknowledge that our target function requires a ground-truth, motion-free scan at optimization time which is an unrealistic requirement for practical applications. In this work, however, the MSE target function serves as the most basic objective which lets us study the performance of different optimization algorithms without any influencing effects of a sub-optimal target function itself. In future studies, the MSE objective should still be replaced by an alternative image quality criterion. Additionally, future work might investigate the generalizability of our results to the 3D cone-beam case, but we are confident that the optimization speed up is even more pronounced in that computationally more expensive setting.
Ultimately, we conclude that gradient-based optimization algorithms are a viable alternative for the studied geometry optimization problem because they accelerate the optimization without sacrificing capture range or robustness to the number of free parameters. 

\section*{Acknowledgements}
The research leading to these results has received funding from the European Research Council under the European Union’s Horizon 2020 research and innovation program (ERC Grant No. 810316).

\printbibliography

\end{document}